\documentclass[onecolumn,a4paper,11pt]{article}

 \usepackage{graphicx}

\usepackage{amssymb}
\usepackage{amsthm,amsmath,amsfonts}
\usepackage{cite}

\date{}

\begin{document}


\title{\begin{flushright}
     {\small IPPP-13-19\\
       DCPT-13-38\\}
       \end{flushright}
       \vskip 20pt
A database for quarkonium and open heavy-flavour production in hadronic collisions with HepData}

\maketitle

\begin{center} 
A.~Andronic$^{a}$, F.~Arl\'eo$^{b}$, R.~Arnaldi$^{c}$, N.~Bastid$^{d}$, G.~Batigne$^{e}$, S.~B\'en\'e$^{d}$, G.~E.~Bruno$^{f}$, J.~Castillo$^{g}$, P.~Crochet$^{d}$, E.~G.~Ferreiro$^{i}$, R.~Granier de Cassagnac$^{j}$, C.~Hadjidakis$^{h}$, G.~Mart\'{\i}nez~Garc\'{\i}a$^{e}$, S.~Masciocchi$^{a}$, S.~Porteboeuf-Houssais$^{d}$\footnote{contact ReteQuarkonii: sarah@clermont.in2p3.fr}, F.~Prino$^{c}$, A.~Rakotozafindrabe$^{g}$, E.~Scomparin$^{c}$ \\
\textbf{for the ReteQuarkonii network} \\

\smallskip
\textbf{and}\\
\smallskip

M. R. Whalley$^{k}$\footnote{contact HepData: hepdata@projects.hepforge.org}\\
\textbf{for the HepData collaboration}.\\

\smallskip
\begin{footnotesize}
\textit{$^{a}$ Research Division and ExtreMe Matter Institute EMMI, GSI Helmholtzzentrum f\"ur Schwerionenforschung, Darmstadt, Germany\\
$^{b}$Laboratoire d'Annecy-le-Vieux de Physique Th\'eorique (LAPTH), UMR5108, Universit\'e de Savoie, CNRS, BP 110, 74941 Annecy-le-Vieux cedex, France\\
$^{c}$Sezione INFN, Turin, Italy\\
$^{d}$Laboratoire de Physique Corpusculaire (LPC), Clermont Universit\'e,Universit\'e Blaise Pascal, CNRS-IN2P3, Clermont-Ferrand, France\\
$^{e}$SUBATECH, Ecole des Mines de Nantes, Universit\'e de Nantes, CNRS-IN2P3, Nantes, France\\
$^{f}$Dipartimento Interateneo di Fisica ``M. Merlin'' and Sezione INFN, Bari, Italy\\
$^{g}$Commissariat \`a l'Energie Atomique, IRFU, Saclay, France\\
$^{h}$Institut de Physique Nucl\'eaire d'Orsay (IPNO), Universit\'e Paris-Sud, CNRS-IN2P3, Orsay, France\\
$^{i}$Departamento de F\'isica de Part\'iculas and IGFAE, Universidad de Santiago de Compostela, E-15782 Santiago de Compostela, Spain\\
$^{j}$LLR, \'Ecole polytechnique, CNRS, F-91128 Palaiseau, France\\
$^{k}$Institute of Particle Physics Phenomenology, Durham University, UK}
\end{footnotesize}

\end{center}
%




\begin{abstract}
We report on the creation of a database for quarkonium and open heavy-flavour production in hadronic collisions. This database, made as a collaboration between HepData and the ReteQuarkonii network of the integrating activity I3HP2 of the 7th Framework Programme, provides an up-to-date review on quarkonia and open heavy-flavour existing data. We first present the physics motivation for this project, which is connected to the aim of the ReteQuarkonii network, studies of open heavy-flavour hadrons  and quarkonia in nucleus-nucleus collisions. Then we give a general overview of the database and describe the HepData database for particle physics, which is the framework of the quarkonia database. Finally we describe the functionalities of the database with as example the comparison of the production cross section for the  J/$\psi$ meson at different energies.

\end{abstract}


\section{Motivation for a Database on quarkonium and open heavy-flavour physics}
\label{motivation}

Quarkonia are bound states of $Q\bar{Q}$ pairs, where $Q$  is a heavy quark, either a charm quark ($c$) or a beauty quark ($b$). The first quarkonium state discovered was the J/$\psi$ particle  ($c\bar{c}$) \cite{Aubert,Augustin}. The quarkonium production mechanism is far from being understood and various models such as the Colour Singlet, nonrelativistic QCD approach (NRQCD) and the Colour Evaporation Model aim to explain how a heavy resonance state can be produced in hard processes, see \cite{Brambilla_2} and references therein.

\bigskip
The study of quarkonia is fundamental for the understanding of the quark-gluon plasma (QGP), a deconfined state of matter produced in ultra-relativistic heavy-ion collisions. In 1986 Matsui and Satz predicted an anomalous suppression of the J/$\psi$ particle  in QGP produced in central heavy-ion collisions \cite{matsui}. A normal suppression is observed in heavy-ion collisions without QGP formation due to the presence of nuclear matter (cold nuclear matter effects). If a QGP is formed, due to colour screening in the hot medium, the J/$\psi$ should be dissociated (hot nuclear matter effects). In addition to this suppression mechanism, theoretical predictions based on recombination models account for an enhancement of J/$\psi$ production due to regeneration in the medium \cite{Svetitsky,Thews} or at the phase space boundary \cite{BraunMunzinger,Andronic, Andronic_2, Zhao}. Heavier states are also of great interest and their binding energies being different, their dissociation temperature should be different too. Theoretical approaches predict a sequential dissociation of quarkonium states, depending on the temperature of the produced medium \cite{Karsch}. 
  
\bigskip
In addition to the study of bound states, open heavy-flavour production (D and B mesons) can also probe hot nuclear matter. Heavy quarks, being produced at initial stage of the collision via hard scattering, interact with the formed QGP. The study of heavy-quark energy loss into the medium gives information on the nature and properties of the QGP (path length, density). Based on QCD, radiative energy loss of quarks should be lower than that of gluons due to the dead cone effect (reduction of in-medium heavy quark energy loss) \cite{Dokshitzer, Djordjevic, Zhang, Armesto}. This effect can be balanced by other mechanisms such as collisional energy loss \cite{Mustafa, Wicks}, in-medium fragmentation, recombination, coalescence \cite{Adil, Greco, vanHees} and initial state effects \cite{Kharzeev,Armesto_2}. 

\bigskip
More details on the production mechanism of heavy-flavour bound states, open heavy-flavours, and their interaction with nuclear medium, either cold or hot can be found in \cite{Vogt,Brambilla,Maltoni,Lansberg,Arleo,Frawley,Rapp,Lansberg_2,Linnyk,Lansberg_3,Kluberg,Rapp_2,Faccioli,Andronic,Brambilla_2}. These effects are extensively studied at the LHC.

\bigskip
For the study of quarkonium and heavy-flavour production in heavy-ion and proton-proton collisions one needs to measure and compare spectra at different energies and from different colliding systems. To disentangle the anomalous from the normal suppression, it is necessary to compare observables in AA collisions with results from pp collisions and pA collisions at the same center of mass energy. 

\bigskip
It is then necessary to have a complete overview of all existing data. This is the motivation for the creation of a database that contains all published results on quarkonia and open heavy-flavours in hadronic collisions. The need for the creation of such a database was pointed out by members of the ReteQuarkonii Network  \cite{retequarkonii}, itself focused on heavy-ion physics. In addition, quarkonia and open heavy-flavours are studied for other physics goals in particle physics and therefore all existing data from hadronic collisions need to be included in the database. This work is done in collaboration with HepData, the Durham high energy physics (HEP) Database Project. Quarkonia related references are included in the database and a dedicated webpage has been created as a review of quarkonium physics where all data are directly accessible \cite{Quarkonia_webpage}.

\section{Overview of the Database}

When dealing with data and databases, the most challenging questions are related to the longevity of the data storage, their accessibility over time and their easy access via the web. This is why this project was done in collaboration with HepData, a well established database in HEP \cite{HEPDATA_site}. In this section, we briefly present HepData and its role in the quarkonia database. Then we present the Quarkonium Review Webpage.   

\subsection{HepData}

HepData's ``reaction database'' is a repository of data from mainly particle physics with some content from nuclear physics. It has been hosted at Durham university since the 1970s with its content based on published data. Data records are stored by publication with  data from approximately 8000 archived papers. More details can be found in \cite{HepData_guide,HepData_2006}. HepData has recently implemented a new software framework based on modern database and programming language technologies, as well as quality tools for the web interface  \cite{HepData_reloaded}.

\begin{figure}[!htp]
	\centering
	\includegraphics[width=1.\columnwidth]{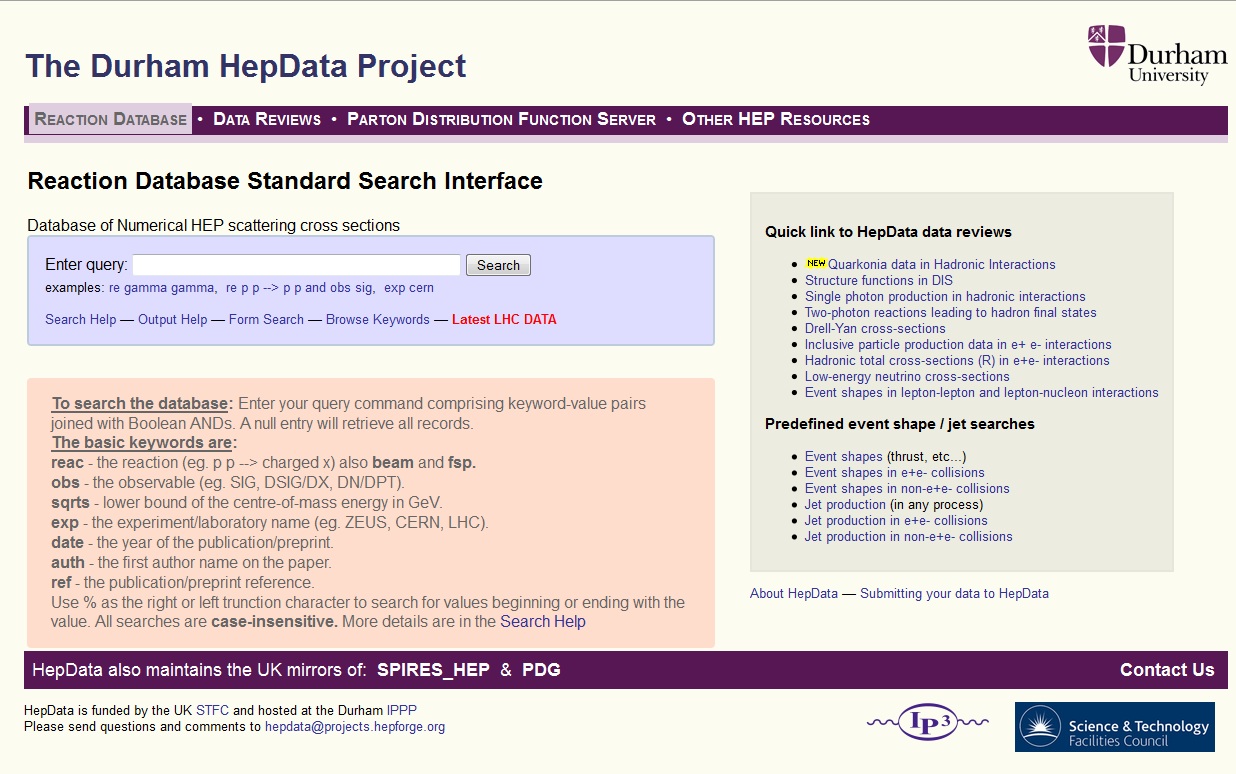}
	\caption{Screenshot of the HepData main webpage \cite{HEPDATA_site}.}
	\label{HepData}
\end{figure}

\begin{figure}[!htp]
	\centering
	\includegraphics[width=1.\columnwidth]{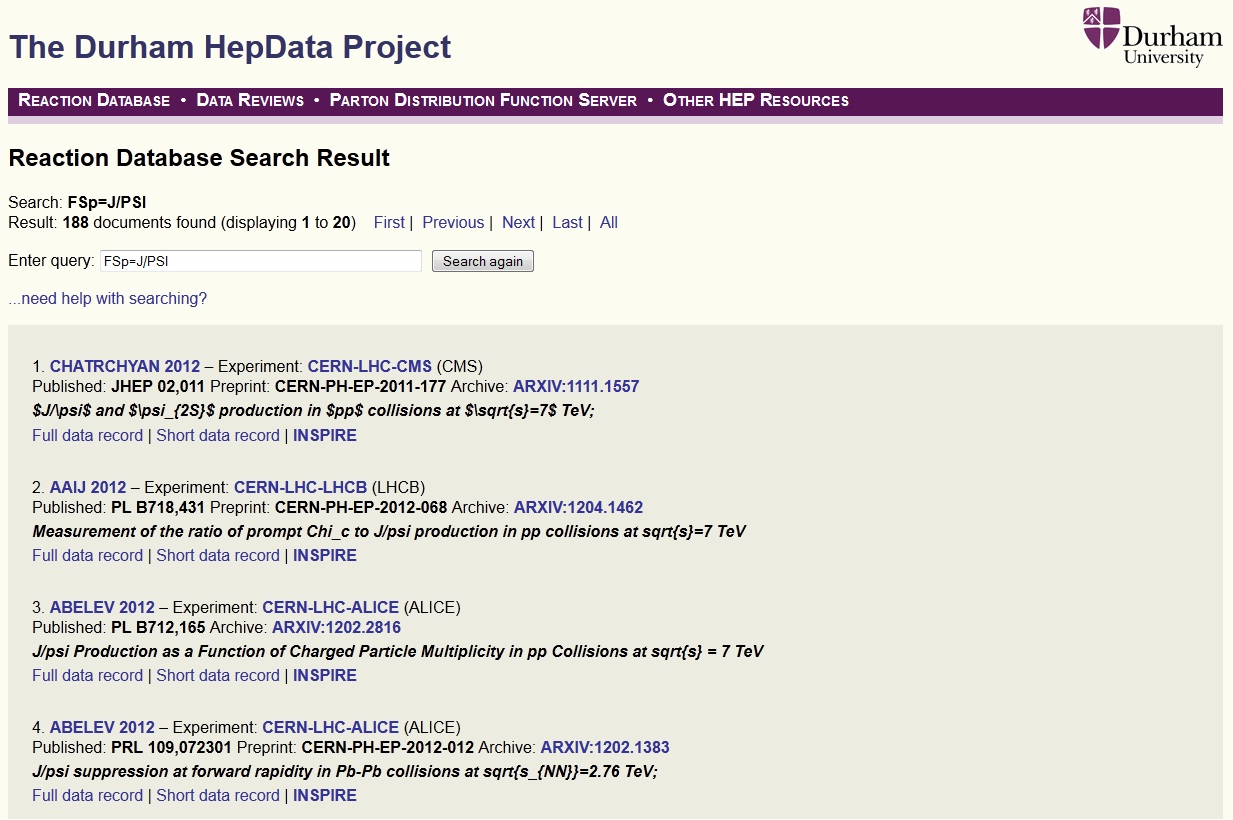}
	\caption{Screenshot of research of records with a J/$\psi$ in the final state.}
	\label{Search_JPSI}
\end{figure}

\bigskip
Figure \ref{HepData} shows a screenshot of the HepData webpage that is accessible via \cite{HEPDATA_site}. On the left hand-side the query form is blank. Users can  search using keywords, such as ``FSP=J/PSI'' to retrieve the records (188 publications at present) with a J/$\psi$ in the final state. The results of this search is shown in Fig. \ref{Search_JPSI},  where only the first four records out of 188 are visible on the screenshot. This also illustrates that latest LHC data are included in HepData.

\bigskip
All the data discussed in this document are included in the traditional HepData database and therefore are accessible via direct search with the query form. 
On the right-hand side of Fig. \ref{HepData} one can see 9 links to data reviews, focused on a specific subject. The Quarkonium Review Webpage is the latest one and will be discussed in the next section.

\subsection{Quarkonium Review Webpage}

\begin{figure}[!htp]
	\centering
	\includegraphics[width=1.\columnwidth]{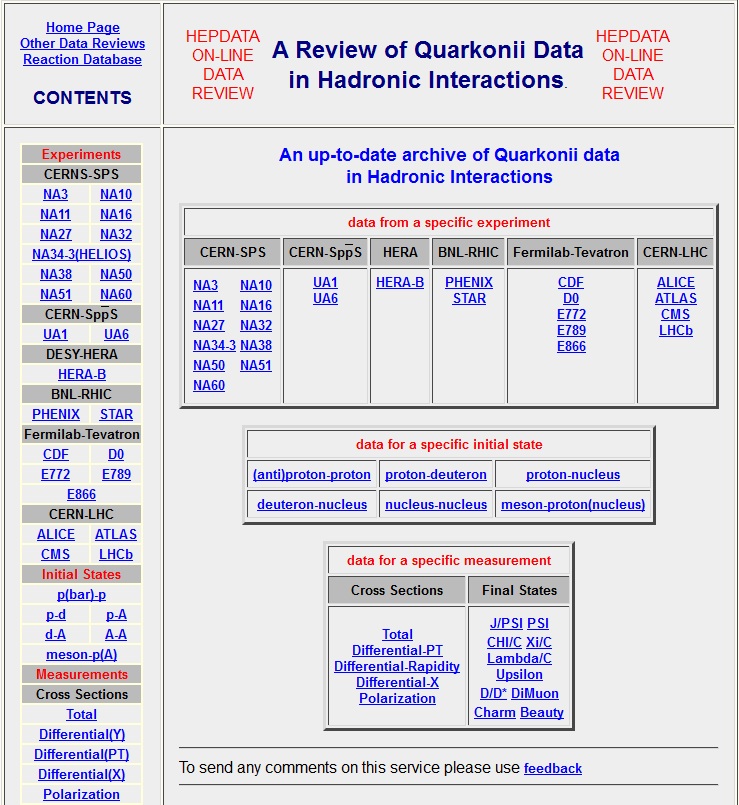}
	\caption{Screenshot of Quarkonium Review Webpage\cite{Quarkonia_webpage}.}
	\label{Quarkonia_webpage}
\end{figure}

HepData offers the possibility to create reviews on dedicated subjects and thus the Quarkonium Review Webpage was created to present all related results in a clear overview. This review contains 185 references at present, and is being updated with new results from the LHC and other experiments. Even though the physics motivation initially came from heavy-ion physics, the database contains data from all experiments which studied quarkonia and open heavy-flavours in hadronic collisions, including data from particle physics experiments. In total, 6 facilities and 25 experiments are considered. Figure \ref{Quarkonia_webpage} shows a screenshot of the Quarkonium Review Webpage, which is accessible on the HepData webpage or directly via\cite{Quarkonia_webpage}.

\bigskip
On the web page, data are first grouped by accelerator facilities and experiments (``data from a specific experiment'', link in red in Fig. \ref{Quarkonia_webpage}). This includes data from the following nuclear and particle physics experiments:

\begin{itemize}

\item CERN-SPS (61 references),  fixed-target experiments with beam energies from 120 GeV to 450 GeV: NA3 \cite{NA3_1,NA3_2,NA3_3,NA3_4,NA3_5 }, NA10\cite{NA10_1,NA10_2,NA10_3,NA10_4}, NA11\cite{NA11_1, NA11_2}, NA16 \cite{NA16_1, NA16_2, NA16_3, NA16_4}, NA27 \cite{NA27_1, NA27_2, NA27_3, NA27_4, NA27_5, NA27_6, NA27_7, NA27_8}, NA32 \cite{NA32_1, NA32_2, NA32_3, NA32_4, NA32_5, NA32_6, NA32_7}, NA34-3 \cite{NA34-3_1}, NA38 \cite{NA38_1,NA38_2,NA38_3,NA38_4,NA38_5,NA38_6,NA38_7,NA38_8,NA38_9,NA38_10,NA38_11}, NA50 \cite{NA50_1,NA50_2,NA50_3,NA50_4,NA50_5,NA50_6,NA50_7,NA50_8,NA50_9,NA50_10,NA50_11,NA50_12,NA50_13,NA50_14,NA50_15,NA50_16}, NA51 \cite{NA51_1}, NA60 \cite{NA60_1,NA60_2};
\item FERMILAB (14 references), fixed-target experiments with beam energies of 800 GeV: E772 \cite{E772_1,E772_2,E772_3,E772_4}, E789 \cite{E789_1,E789_2,E789_3,E789_4,E789_5,E789_6}, E866 \cite{E866_1,E866_2,E866_3,E866_4};
\item HERA (12 references), fixed-target experiments with beam energies of 920 GeV: HERA-b \cite{HERA_1,HERA_2,HERA_3,HERA_4,HERA_5,HERA_6,HERA_7,HERA_8,HERA_9,HERA_10,HERA_11,HERA_12};
\item BNL-RHIC  (20 references), collider experiments with $\sqrt{s}=200$ GeV: PHENIX \cite{PHENIX_1,PHENIX_2,PHENIX_3,PHENIX_4,PHENIX_5,PHENIX_6,PHENIX_7,PHENIX_8, PHENIX_O_1,PHENIX_O_2, PHENIX_O_3, PHENIX_O_4, PHENIX_O_5,PHENIX_O_6,PHENIX_O_7,PHENIX_O_8} , STAR \cite{STAR_1, STAR_O_1, STAR_O_2, STAR_O_4};
\item CERN SP$\bar{\rm{P}}$S (7 references), collider experiments with $\sqrt{s}=540-630$ GeV: UA1  \cite{UA1_1, UA1_2, UA1_3, UA1_4, UA1_5, UA1_6}, UA6 \cite{UA6_1};
\item Fermilab-Tevatron (33 references), collider experiments with $\sqrt{s}=1.8-1.96$ TeV: CDF \cite{CDF_1,CDF_2,CDF_3,CDF_4,CDF_5,CDF_6,CDF_7,CDF_8,CDF_9,CDF_10,CDF_11,CDF_12,CDF_13
,CDF_14,CDF_15,CDF_16,CDF_17,CDF_18,CDF_19,CDF_20,CDF_21,CDF_22,CDF_23,CDF_24}, D0 \cite{D0_1,D0_2,D0_3,D0_4,D0_5,D0_6,D0_7,D0_8,D0_9};
\item CERN-LHC (38 references), collider experiments with $\sqrt{s}=2.76-7$ TeV:  ALICE, ATLAS, CMS, LHCb  \cite{ALICE_1,ALICE_2,ALICE_3,ALICE_4,ALICE_5,ALICE_6,ALICE_7,ALICE_8}, ATLAS\cite{ATLAS_1,ATLAS_2,ATLAS_3,ATLAS_4,ATLAS_5,ATLAS_6,ATLAS_7,ATLAS_8}, CMS \cite{CMS_1, CMS_2, CMS_3, CMS_4, CMS_5, CMS_6, CMS_7, CMS_8, CMS_9, CMS_10, CMS_11, CMS_12, CMS_13},  LHCb \cite{LHCb_1,LHCb_2,LHCb_3,LHCb_4,LHCb_5,LHCb_6,LHCb_7,LHCb_8,LHCb_9}.
\end{itemize}

\bigskip
Next, in order to make easier the search of a specific results, data are sorted out by initial state (``data for a specific initial state'', link in red in Fig. \ref{Quarkonia_webpage}):
\begin{itemize}
\item proton-(anti)proton, proton-deuteron, deuteron-deuteron;
\item proton-nucleus;
\item deuteron-nucleus and nucleus-nucleus;
\item meson-nucleus.
\end{itemize}

\bigskip
Finally, data are sorted out according to specific measurements and observables (``data for a specific measurement'', link in red in Fig. \ref{Quarkonia_webpage}): 
\begin{itemize}
\item integrated cross sections, differential cross sections versus $p_{\rm T}$,  rapidity and  $x$;
\item polarization;
\end{itemize}
 and by final states particles:
\begin{itemize}
\item J/$\psi$, $\psi$, $\chi$(c) and $\Upsilon$;
\item $\rm{D}$, $\rm{D}^*$, di-muon, charm and beauty.
\end{itemize}

\begin{figure}[!htp]
	\centering
	\includegraphics[width=1.\columnwidth]{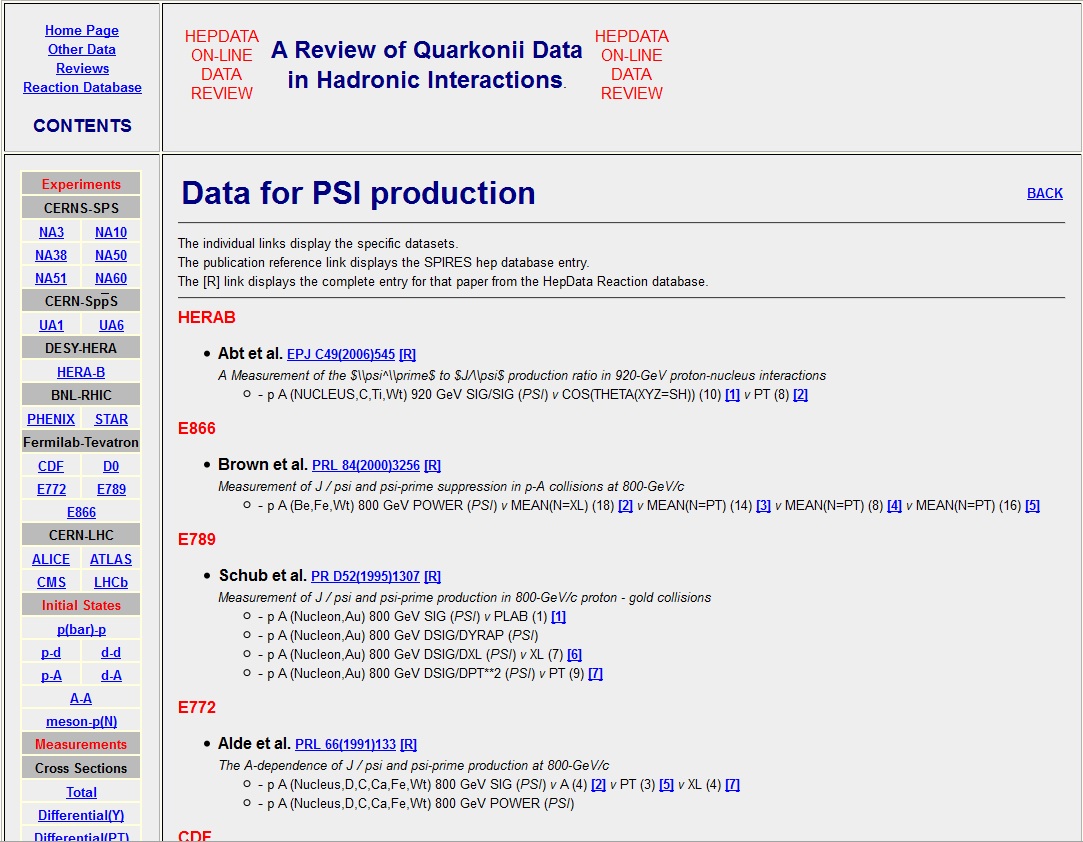}
	\caption{Screenshot of Quarkonium Review Webpage for the production of a J/$\psi$ particle in the final state.}
	\label{JPSI_webpage}
\end{figure}

\bigskip
When selecting data in a specific search, as for example J/$\psi$ in Fig. \ref{Quarkonia_webpage}, all related papers are listed, as shown in Fig. \ref{JPSI_webpage}. In blue, there is a link to inSPIRE \cite{Inspires_webpage} where a PDF version of the publication can usually be obtained. The ``[R]'' link points to the full HepData record where all available plots in the paper can be found and the traditional HepData machinery can be used to visualize data tables and plot figures. This aspect will be discussed in the next section.   

\section{Use of the database}
\label{use}

\begin{figure}[!htp]
	\centering
	\includegraphics[width=0.9 \columnwidth]{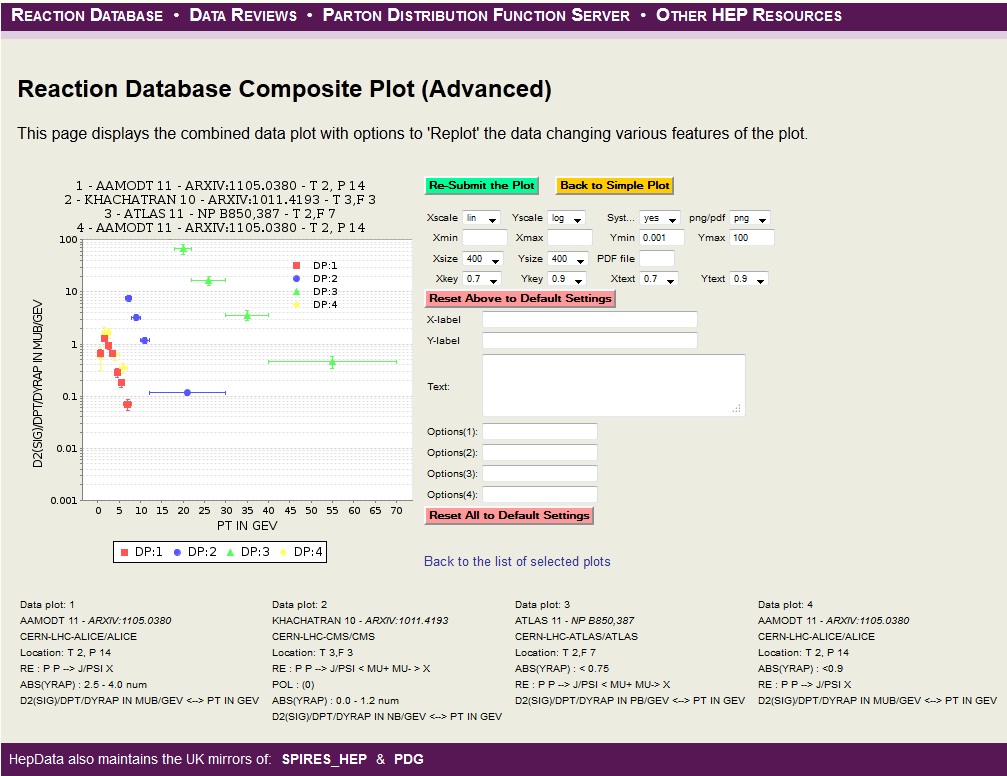}
	\caption{Screenshot of the advanced graphic plotting option from the HepData plotting tool.}
	\label{graphic_interafce}
\end{figure}

To quickly and easily compare data sets, one can use the graphical HepData tool available online. For each column of a data table the link ``select plot'' allows the user to select several data tables to be displayed in the same plot. Each data set is referenced by a number. Numbers are set in order of selection (the number one is attributed to first table). Graphics can then be customized. The advanced graphic interface (see Fig. \ref{graphic_interafce}) is user friendly with predefined fields where the user can easily select options and features such as:
\begin{itemize}
\item the size and aspect of the plot can be changed (Xsize,Ysize);
\item axes can be set linear or logarithmic (Xscale, Yscale);
\item axes range can be fixed by filling the xmin, xmax, ymin, ymax boxes;
\item axes can be re-labeled (X-label, Y-label);
\item a text box allows the addition of a  title or comments to the plot (Text);
\item the position of the labels and text can be adjusted (Xsize, Ysize);
\item a factor can be applied to a data set with the scale command, for example ``scale=5'' (``5'' is the factor to be applied to all data points in that particular set). This command is essential to compare data sets provided with different units, for example one in nanobarn and the other in microbarn (Option(n) for data set n);
\item colour and icon type can be modified with the use of colour name and shape in the relevant option box (black, pink, cyan, green, square, diamond, triangle, filled, etc) (Option(n) for data set n). All commands in the option boxes are comma separated.

\end{itemize}
All data comparisons presented in this section were made online using the HepData graphical tool. Thus, they all can be easily reproduced by anyone. 
The HepData community is continuously working to improve the graphical tool.

\begin{figure}[h!]
   \begin{minipage}[b]{0.48\linewidth}
\includegraphics[width=1. \columnwidth]{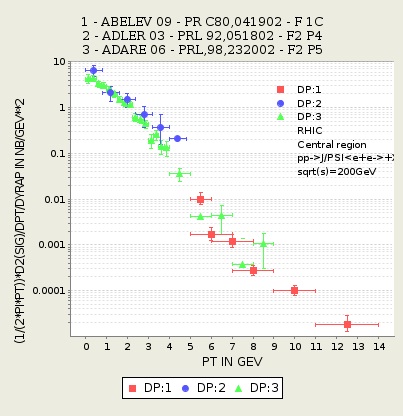}
   \end{minipage}\hfill
   \begin{minipage}[b]{0.48\linewidth}  
\includegraphics[width=1.\columnwidth]{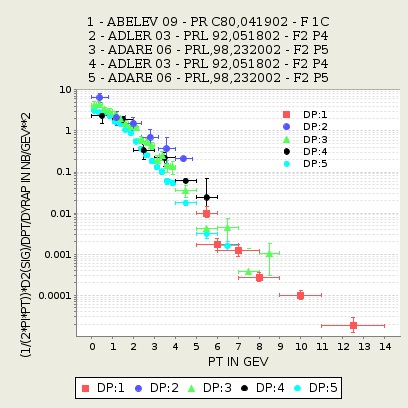}
   \end{minipage}
   \caption{
J/$\psi$ $p_{\rm T}$ spectra at RHIC energies ($\sqrt{s}=200$ GeV).  Left: central rapidity region. Right: central and forward rapidity regions. 
}
   \label{RHIC}
\end{figure}

\bigskip
Various comparison plots can be made to illustrate sample data records in the quarkonia database. Here we will focus on J/$\psi$ production in pp collisions at energies from RHIC to LHC. Figure \ref{RHIC} shows, on the left plot, results in the central rapidity region. STAR data are obtained with e$^+$e$^-$ pairs, in the region $|\eta |<0.5$ (red squares,1) and PHENIX data with e$^+$e$^-$ pairs in  $|y|<0.35$ (blue circles, 2 and green triangles, 3). Only statistical uncertainties are shown, systematic uncertainties are available in data tables. One can see a good agreement between the two PHENIX measurements and the STAR one, in complementary $p_{\rm T}$ ranges. On the right plot, PHENIX results with $\mu^+ \mu^-$ pairs in the region $-2.2<|y|<-1.2$ are added (black circles, 4 and light blue circles, 5).

\begin{figure}[!htp]
	\centering
	\includegraphics[width=0.5 \columnwidth]{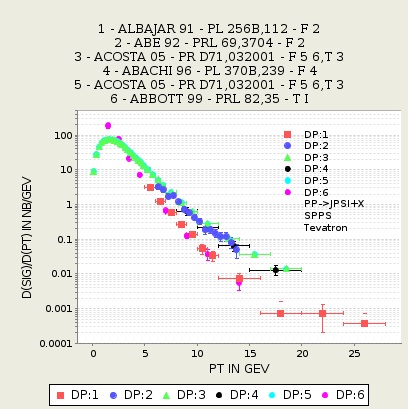}
	\caption{
J/$\psi$ $p_{\rm T}$ spectra in p$\bar{\rm{p}}$ collisions at SP$\bar{\rm{P}}$S energies ($\sqrt{s}=630$ GeV) with UA1 experiment and at Tevatron energies ($\sqrt{s}=1.8-1.96$ TeV) by CDF and D0 experiments. 
 }
	\label{pt}
\end{figure}

\bigskip
Figure \ref{pt} shows a similar comparison for results from UA1, CDF and D$0$. 1 (red squares) are UA1 results at $\sqrt{s}=630$ GeV. Labels 2, 3, 5 (blue circles, green triangles and light blue circles) are CDF results at $\sqrt{s}=1.8-1.96$ TeV. Labels 4, 6 (black circles and pink circles) are D0 at $\sqrt{s}=1.8-1.96$ TeV. Also here, only statistical uncertainties are shown, systematic uncertainties are available in the data table. In this plot, one can see the $p_{\rm T}$ spectrum of the J/$\psi$ production becoming harder with increasing energies.

\begin{figure}[h!]
   \begin{minipage}[b]{0.48\linewidth}
 \includegraphics[width=1. \columnwidth]{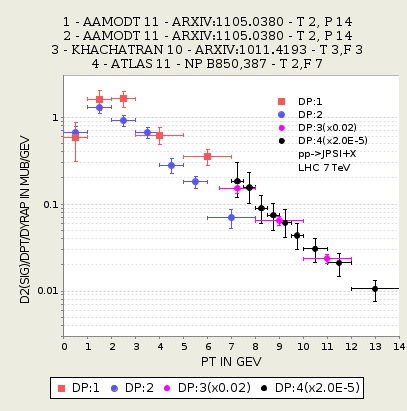}
   \end{minipage}\hfill
   \begin{minipage}[b]{0.48\linewidth}  
\includegraphics[width=1. \columnwidth]{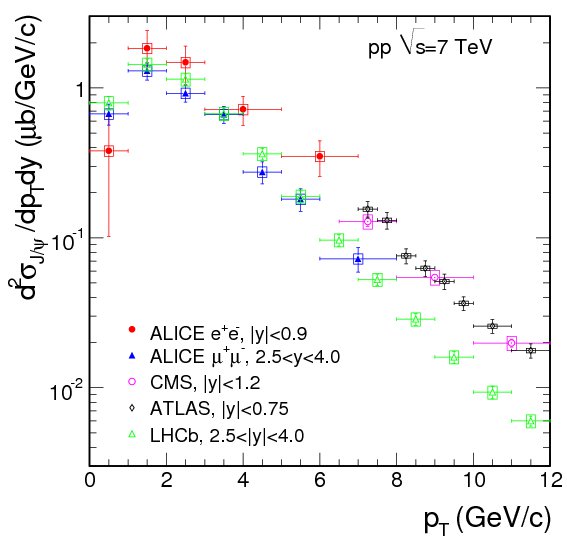}
   \end{minipage}
   \caption{J/$\psi$ $p_{\rm T}$ spectra in pp collisions at $\sqrt{s}=7$ TeV  at mid-rapidity (ALICE, CMS and ATLAS) and forward rapidity (ALICE and LHCb). Left: plot from HepData. Right: plot from ALICE publication \cite{ALICE_1}.}
   \label{LHC}
\end{figure}

\bigskip
Figure \ref{LHC} displays $p_{\rm T}$ distributions of J/$\psi$ produced in pp collisions at $\sqrt{s}=7$ TeV, at mid-rapidity (ALICE, CMS and ATLAS) and forward rapidity (ALICE and LHCb). The right plot was published by the ALICE collaboration \cite{ALICE_1}. On the left, the same plot was reproduced with the HepData graphical tool with statistical uncertainties only. LHCb data are missing since they deal with direct J/$\psi$ production and not inclusive production.

\bigskip
This exercise shows the possibility of using the Quarkonium Review Webpage with HepData graphical tools, to plot selected data sets. This tool is very useful to have a first quick comparison of data results from different experiments.

\section{Conclusion}
We have reported on the creation of a database dedicated to quarkonia and open heavy-flavour physics in hadronic collisions. The need for this work was highlighted by the ReteQuarkonii network members and has been done in collaboration with the Durham HepData project whose ``reaction database'' provides the framework for the quarkonium and open heavy-flavour review webpage. We have included data from 25  experiments in this review from SPS to LHC energies and we have demonstrated the possibility of performing quick data comparison online with the HepData graphical tool. HepData is continuously updated with new data and the Quarkonium Review Webpage will also include the new results. We encourage the use of this database and to report anomalies to us\footnote{contact ReteQuarkonii: sarah@clermont.in2p3.fr, \\ contact HepData:  hepdata@projects.hepforge.org}.  An extension of this database could be foreseen with the inclusion of data from electron colliders.

\section*{Acknowledgments}

This publication has been supported by the European Commission under the Integrating Activity HadronPhysics2 (Grant agreement n. 283286) of the 7th Framework Programme. Sarah Porteboeuf-Houssais would like to thank the ReteQuarkonii Network and the I3HP2 European Program which provided the grant supporting this work and the STFC(UK) is thanked for its continuous support of HepData. 

\bibliographystyle{ieeetr}
\bibliography{ReteQuarkonii_bibliography}

\begin{thebibliography}{100}

\bibitem{Aubert}
J.~J. Aubert {\em et~al.}, ``{Discovery of the New Particle J},'' {\em Nucl.
  Phys.}, vol.~B 89, p.~1, 1975.

\bibitem{Augustin}
J.~E. Augustin {\em et~al.}, ``{Discovery of a Narrow Resonance in e+ e-
  Annihilation},'' {\em Phys. Rev. Lett.}, vol.~33, pp.~1406--1408, 1974.

\bibitem{Brambilla_2}
N.~Brambilla {\em et~al.}, ``{Heavy quarkonium: progress, puzzles, and
  opportunities},'' {\em Eur. Phys. J.}, vol.~C 71, p.~1534, 2011.

\bibitem{matsui}
T.~Matsui and H.~Satz, ``{J/$\psi$ Suppression by Quark-Gluon Plasma
  Formation},'' {\em Phys. Lett.}, vol.~B 178, p.~416, 1986.

\bibitem{Svetitsky}
B.~Svetitsky, ``{Diffusion of charmed quarks in the quark-gluon plasma},'' {\em
  Phys. Rev.}, vol.~D 37, pp.~2484--2491, 1988.

\bibitem{Thews}
R.~L. Thews, M.~Schroedter, and J.~Rafelski, ``{Enhanced J/$\psi$ production in
  deconfined quark matter},'' {\em Phys. Rev.}, vol.~C 63, p.~054905, 2001.

\bibitem{BraunMunzinger}
P.~Braun-Munzinger and J.~Stachel, ``{(Non)thermal aspects of charmonium
  production and a new look at J/$\psi$ suppression},'' {\em Phys. Lett.},
  vol.~B 490, pp.~196--202, 2000.

\bibitem{Andronic}
A.~Andronic, F.~Beutler, P.~Braun-Munzinger, K.~Redlich, and J.~Stachel,
  ``{Statistical hadronization of heavy flavor quarks in elementary collisions:
  successes and failures},'' {\em Phys. Lett.}, vol.~B 678, pp.~350--354, 2009.

\bibitem{Andronic_2}
A.~Andronic, P.~Braun-Munzinger, K.~Redlich, and J.~Stachel, ``{Evidence for
  charmonium generation at the phase boundary in ultra-relativistic nuclear
  collisions},'' {\em Phys. Lett.}, vol.~B 652, pp.~259--261, 2007.

\bibitem{Zhao}
X.~Zhao and R.~Rapp, ``{Transverse Momentum Spectra of J/$\psi$ in Heavy-Ion
  Collisions},'' {\em Phys. Lett.}, vol.~B 664, pp.~253--257, 2008.

\bibitem{Karsch}
F.~Karsch, D.~Kharzeev, and H.~Satz, ``{Sequential charmonium dissociation},''
  {\em Phys. Lett.}, vol.~B 637, pp.~75--80, 2006.

\bibitem{Dokshitzer}
Y.~L. Dokshitzer and D.~Kharzeev, ``{Heavy quark colorimetry of QCD matter},''
  {\em Phys. Lett.}, vol.~B 519, pp.~199--206, 2001.

\bibitem{Djordjevic}
M.~Djordjevic and M.~Gyulassy, ``{Heavy quark radiative energy loss in QCD
  matter},'' {\em Nucl. Phys.}, vol.~A 733, pp.~265--298, 2004.

\bibitem{Zhang}
B.-W. Zhang, E.~Wang, and X.-N. Wang, ``{Heavy quark energy loss in nuclear
  medium},'' {\em Phys. Rev. Lett.}, vol.~93, p.~072301, 2004.

\bibitem{Armesto}
N.~Armesto, C.~A. Salgado, and U.~A. Wiedemann, ``{Medium induced gluon
  radiation off massive quarks fills the dead cone},'' {\em Phys. Rev.}, vol.~D
  69, p.~114003, 2004.

\bibitem{Mustafa}
M.~G. Mustafa, ``{Energy loss of charm quarks in the quark-gluon plasma:
  Collisional versus radiative},'' {\em Phys. Rev.}, vol.~C 72, p.~014905,
  2005.

\bibitem{Wicks}
S.~Wicks, W.~Horowitz, M.~Djordjevic, and M.~Gyulassy, ``{Heavy quark jet
  quenching with collisional plus radiative energy loss and path length
  fluctuations},'' {\em Nucl. Phys.}, vol.~A 783, pp.~493--496, 2007.

\bibitem{Adil}
A.~Adil and I.~Vitev, ``{Collisional dissociation of heavy mesons in dense QCD
  matter},'' {\em Phys. Lett.}, vol.~B 649, pp.~139--146, 2007.

\bibitem{Greco}
V.~Greco, C.~Ko, and R.~Rapp, ``{Quark coalescence for charmed mesons in
  ultrarelativistic heavy ion collisions},'' {\em Phys. Lett.}, vol.~B 595,
  pp.~202--208, 2004.

\bibitem{vanHees}
H.~van Hees, V.~Greco, and R.~Rapp, ``{Heavy-quark probes of the quark-gluon
  plasma at RHIC},'' {\em Phys. Rev.}, vol.~C 73, p.~034913, 2006.

\bibitem{Kharzeev}
D.~Kharzeev, E.~Levin, and L.~McLerran, ``{Parton saturation and N(part)
  scaling of semihard processes in QCD},'' {\em Phys. Lett.}, vol.~B 561,
  pp.~93--101, 2003.

\bibitem{Armesto_2}
N.~Armesto, ``{Nuclear shadowing},'' {\em J. Phys.}, vol.~G 32, pp.~R367--R394,
  2006.

\bibitem{Vogt}
R.~Vogt, ``{J/$\psi$ production and suppression},'' {\em Phys. Rept.},
  vol.~310, pp.~197--260, 1999.

\bibitem{Brambilla}
N.~Brambilla {\em et~al.}, ``{Heavy quarkonium physics},'' 2004.
\newblock arXiv: hep-ph/0412158.

\bibitem{Maltoni}
F.~Maltoni {\em et~al.}, ``{Analysis of charmonium production at fixed-target
  experiments in the NRQCD approach},'' {\em Phys. Lett.}, vol.~B 638,
  pp.~202--208, 2006.

\bibitem{Lansberg}
J.~P. Lansberg, ``{J/$\psi$, $\psi$' and $\Upsilon$ production at hadron
  colliders: A Review},'' {\em Int. J. Mod. Phys.}, vol.~A 21, pp.~3857--3916,
  2006.

\bibitem{Arleo}
F.~Arleo and V.-N. Tram, ``{A systematic study of J/$\psi$ suppression in cold
  nuclear matter},'' {\em Eur. Phys. J.}, vol.~C 55, pp.~449--461, 2008.

\bibitem{Frawley}
A.~D. Frawley, T.~Ullrich, and R.~Vogt, ``{Heavy flavor in heavy-ion collisions
  at RHIC and RHIC II},'' {\em Phys. Rept.}, vol.~462, pp.~125--175, 2008.

\bibitem{Rapp}
R.~Rapp, D.~Blaschke, and P.~Crochet, ``{Charmonium and bottomonium production
  in heavy-ion collisions},'' {\em Prog. Part. Nucl. Phys.}, vol.~65,
  pp.~209--266, 2010.

\bibitem{Lansberg_2}
J.~Lansberg, ``{On the mechanisms of heavy-quarkonium hadroproduction},'' {\em
  Eur. Phys. J.}, vol.~C 61, pp.~693--703, 2009.

\bibitem{Linnyk}
O.~Linnyk, E.~L. Bratkovskaya, and W.~Cassing, ``{Open and hidden charm in
  proton-nucleus and heavy-ion collisions},'' {\em Int. J. Mod. Phys.}, vol.~E
  17, pp.~1367--1439, 2008.

\bibitem{Lansberg_3}
J.~P. Lansberg, ``{On the mechanisms of heavy-quarkonium hadroproduction},''
  {\em Eur. Phys. J.}, vol.~C 61, pp.~693--703, 2009.

\bibitem{Kluberg}
L.~Kluberg and H.~Satz, ``{Color Deconfinement and Charmonium Production},''
  2009.
\newblock arXiv: hep-ph/0901.383.

\bibitem{Rapp_2}
R.~Rapp and H.~van Hees, ``{Heavy Quarks in the Quark-Gluon Plasma},'' 2009.
\newblock arXiv: hep-ph/0903.1096.

\bibitem{Faccioli}
P.~Faccioli, C.~Lourenco, J.~Seixas, and H.~K. Wohri, ``{J/$\psi$ polarization
  from fixed-target to collider energies},'' {\em Phys. Rev. Lett.}, vol.~102,
  p.~151802, 2009.

\bibitem{retequarkonii}
ReteQuarkonii is a network of the Integrating Activity HadronPhysics2 (Grant
  agreement n. 283286) of the 7th Framework Programme. Twiki web link:
  https://twiki.cern.ch/twiki/bin/view/ReteQuarkonii.

\bibitem{Quarkonia_webpage}
http://hepdata.cedar.ac.uk/review/quarkonii/.

\bibitem{HEPDATA_site}
http://hepdata.cedar.ac.uk/.

\bibitem{HepData_guide}
M.~Whalley, ``{A Guide to using HEPDATA (the Durham/RAL databases) on the World
  Wide Web},''
\newblock DPDG-97-02.

\bibitem{HepData_2006}
A.~Buckley {\em et~al.}, ``{HepData and JetWeb: HEP data archiving and model
  validation},'' 2006.
\newblock arXiv: hep-ph/0605048.

\bibitem{HepData_reloaded}
A.~Buckley and M.~Whalley, ``{HepData reloaded: reinventing the HEP data
  archive},'' {\em PoS}, vol.~ACAT 2010, p.~067, 2010.

\bibitem{NA3_1}
J.~Badier {\em et~al.}, ``{First evidence for $\Upsilon$ production by
  pions},'' {\em Phys. Lett.}, vol.~B 86, p.~98, 1979.

\bibitem{NA3_2}
J.~Badier {\em et~al.}, ``{Evidence for $\psi$-$\psi$ production in $\pi^-$
  interactions at 150-GeV/c and 280-GeV/c},'' {\em Phys. Lett.}, vol.~B 114,
  p.~457, 1982.

\bibitem{NA3_3}
J.~Badier {\em et~al.}, ``{Upperlimits on beauty mesons production in $\pi^-$
  collisions at 280-GeV/c},'' {\em Phys. Lett.}, vol.~B 124, p.~535, 1983.

\bibitem{NA3_4}
J.~Badier {\em et~al.}, ``{Experimental J/$\psi$ Hadronic Production from
  150-GeV/c to 280-GeV/c},'' {\em Z. Phys.}, vol.~C 20, p.~101, 1983.

\bibitem{NA3_5}
J.~Badier {\em et~al.}, ``{$\psi$-$\psi$ production and limits on beauty meson
  production from 400-GeV/c protons},'' {\em Phys. Lett.}, vol.~B 158, p.~85,
  1985.

\bibitem{NA10_1}
S.~Falciano {\em et~al.}, ``{A-dependance of muons pair production in $\pi^-$
  nucleus interactions at 280-GeV/c},'' {\em Phys. Lett.}, vol.~B 104, p.~416,
  1981.

\bibitem{NA10_2}
B.~Betev {\em et~al.}, ``{Differential cross-section of high mass muon pairs
  produced by a 194-GeV/c $\pi^-$ beam on a tungsten target },'' {\em Z.
  Phys.}, vol.~C 28, p.~9, 1985.

\bibitem{NA10_3}
M.~Grossmann-Handschin {\em et~al.}, ``{A high statistics study of $\Upsilon$
  meson production in $\pi^-$W reactions at 286-GeV/c},'' {\em Phys. Lett.},
  vol.~B 179, p.~170, 1986.

\bibitem{NA10_4}
P.~Bordalo {\em et~al.}, ``{Open beauty production in high-energy $\pi^-$
  tungsten interactions},'' {\em Z. Phys.}, vol.~C 39, p.~7, 1988.

\bibitem{NA11_1}
R.~Bailey {\em et~al.}, ``{Measurement of D meson production in 200-GeV $\pi^-$
  Be interactions},'' {\em Z. Phys.}, vol.~C 30, p.~51, 1986.

\bibitem{NA11_2}
R.~Bailey {\em et~al.}, ``{Observation of D$^{*+/-}$ and $\bar{\rm{D}}_0$ /
  D$^{\pm}$ production in high-energy $\pi^-$ Be interactions at the sps},''
  {\em Phys. Lett.}, vol.~B 132, p.~230, 1983.

\bibitem{NA16_1}
M.~Aguilar-Benitez {\em et~al.}, ``{D meson branching ratios and hadronic charm
  production cross-sections},'' {\em Phys. Lett.}, vol.~B 135, p.~237, 1984.

\bibitem{NA16_2}
M.~Aguilar-Benitez {\em et~al.}, ``{Charm D meson production in 360-GeV/c pp
  interactions: comparison with $\pi^-$ p at the same energy},'' {\em Phys.
  Lett.}, vol.~B 123, p.~103, 1983.

\bibitem{NA16_3}
M.~Aguilar-Benitez {\em et~al.}, ``{Charm D meson production in 360-GeV $\pi^-$
  p interactions: evidence for leading quarks},'' {\em Phys. Lett.}, vol.~B
  123, p.~98, 1983.

\bibitem{NA16_4}
M.~Aguilar-Benitez {\em et~al.}, ``{D meson production from 400-GeV/c pp
  interactions},'' {\em Phys. Lett.}, vol.~B 189, p.~476, 1987.

\bibitem{NA27_1}
M.~Aguilar-Benitez {\em et~al.}, ``{Neutral and charged D* production in
  360-GeV/c $\pi^-$ p interactions},'' {\em Phys. Lett.}, vol.~B 169, p.~106,
  1986.

\bibitem{NA27_2}
M.~Aguilar-Benitez {\em et~al.}, ``{Neutral D meson properties in 360-GeV/c
  $\pi^-$ p interactions},'' {\em Phys. Lett.}, vol.~B 146, p.~266, 1984.

\bibitem{NA27_3}
M.~Aguilar-Benitez {\em et~al.}, ``{Charm Hadron Properties in 360-GeV/c
  $\pi^-$ p Interactions},'' {\em Z. Phys.}, vol.~C 31, p.~491, 1986.

\bibitem{NA27_4}
M.~Aguilar-Benitez {\em et~al.}, ``{Inclusive Properties of D Mesons Produced
  in 360-GeV $\pi^-$ p Interactions},'' {\em Phys. Lett.}, vol.~B 161,
  pp.~400--406, 1985.

\bibitem{NA27_5}
M.~Aguilar-Benitez {\em et~al.}, ``{D$\bar{\rm{D}}$ correlations in 360-GeV/c
  $\pi^-$ p interactions},'' {\em Phys. Lett.}, vol.~B 164, p.~404, 1985.

\bibitem{NA27_6}
M.~Aguilar-Benitez {\em et~al.}, ``{$\Lambda$(c) production characteristics in
  proton proton interactions at 400-GeV/c},'' {\em Phys. Lett.}, vol.~B 199,
  p.~462, 1987.

\bibitem{NA27_7}
M.~Aguilar-Benitez {\em et~al.}, ``{D meson production from 400-GeV/c pp
  interactions. evidence for leading diquarks?},'' {\em Phys. Lett.}, vol.~B
  201, p.~176, 1988.

\bibitem{NA27_8}
M.~Aguilar-Benitez {\em et~al.}, ``{Comparative properties of 400-GeV/c proton
  - proton interactions with and without charm production},'' {\em Z. Phys.},
  vol.~C 41, p.~191, 1988.

\bibitem{NA32_1}
S.~Barlag {\em et~al.}, ``{Charmed pair correlations in $\pi^-$ Cu interactions
  at 230- GeV/c},'' {\em Phys. Lett.}, vol.~B 302, pp.~112--118, 1993.

\bibitem{NA32_2}
S.~Barlag {\em et~al.}, ``{Production properties of D$^0$, D$^+$, D$^{*+}$ and
  D$_s^+$ in 230- GeV/c $\pi^-$ and K- Cu interactions},'' {\em Z. Phys.},
  vol.~C 49, pp.~555--562, 1991.

\bibitem{NA32_3}
S.~Barlag {\em et~al.}, ``{Production of the charmed baryon $\Lambda$(c)+ in
  $\pi^-$ Cu and K- Cu interactions at 230-GeV},'' {\em Phys. Lett.}, vol.~B
  247, pp.~113--120, 1990.

\bibitem{NA32_4}
S.~Barlag {\em et~al.}, ``{First measurement of the lifetime of the charmed
  strange baryon $\chi_c^0$},'' {\em Phys. Lett.}, vol.~B 236, p.~495, 1990.

\bibitem{NA32_5}
S.~Barlag {\em et~al.}, ``{Measurement of the mass and lifetime of the charmed
  strange baryon $\chi_c^+$},'' {\em Phys. Lett.}, vol.~B 233, p.~522, 1989.

\bibitem{NA32_6}
S.~Barlag {\em et~al.}, ``{Results on $\Lambda$(c)+, D$_s^+$, D$^0$ and D$^+$
  production properties in 230-GeV/c $\pi^-$ Cu interactions from the NA32
  experiment},''
\newblock Contribution to 24th int. Conf. on High Energy Physics, Munich, West
  Germany, Aug 4-10, 1988.

\bibitem{NA32_7}
R.~Bailey {\em et~al.}, ``{Upper limits for charm production in 150-GeV p Be
  interactions},'' {\em Nucl. Phys.}, vol.~B 239, p.~15, 1984.

\bibitem{NA34-3_1}
A.~L.~S. Angelis {\em et~al.}, ``{Excess of continuum dimuon production at
  masses between threshold and the J/$\psi$ in S-W interactions at 200-
  GeV/c/nucleon},'' {\em Eur. Phys. J.}, vol.~C 13, pp.~433--452, 2000.

\bibitem{NA38_1}
M.~C. Abreu {\em et~al.}, ``{Transverse momentum of J/$\psi$, $\psi$' and mass
  continuum muon pairs produced in S-32 U collisions at 200-GeV/c per
  nucleon},'' {\em Phys. Lett.}, vol.~B 423, pp.~207--212, 1998.

\bibitem{NA38_2}
M.~C. Abreu {\em et~al.}, ``{Charmonia production in 450-GeV/c proton induced
  reactions},'' {\em Phys. Lett.}, vol.~B 444, pp.~516--522, 1998.

\bibitem{NA38_3}
M.~C. Abreu {\em et~al.}, ``{J/$\psi$, $\psi$' and Drell-Yan production in S-U
  interactions at 200-GeV per nucleon},'' {\em Phys. Lett.}, vol.~B 449,
  pp.~128--136, 1999.

\bibitem{NA38_4}
M.~C. Abreu {\em et~al.}, ``{J/$\psi$ and $\psi$' production in p, O and S
  induced reactions at SPS energies},'' {\em Phys. Lett.}, vol.~B 466,
  pp.~408--414, 1999.

\bibitem{NA38_5}
C.~Baglin {\em et~al.}, ``{$\psi$' and J/$\psi$ production in p-W, p-U and S-U
  interactions at 200-GeV/nucleon},'' {\em Phys. Lett.}, vol.~B 345,
  pp.~617--621, 1995.

\bibitem{NA38_6}
M.~C. Abreu {\em et~al.}, ``{Transverse momentum of dimuons production in p-U,
  O-U and S-U collisions at 200-GeV/nucleon},'' {\em Nucl. Phys.}, vol.~A 525,
  pp.~469c--472c, 1991.

\bibitem{NA38_7}
C.~Baglin {\em et~al.}, ``{J/$\psi$ and muon-pair cross-sections in
  proton-nucleus and nucleus-nucleus collisions at 200 GeV per nucleon},'' {\em
  Phys. Lett.}, vol.~B 270, pp.~105--110, 1991.

\bibitem{NA38_8}
C.~Lourenco {\em et~al.}, ``{J/$\psi$, $\psi$' and muon pair production in p-W
  and S-U collisions},'' {\em Nucl. Phys.}, vol.~A 566, pp.~77c--85c, 1994.

\bibitem{NA38_9}
C.~Baglin {\em et~al.}, ``{Transverse momentum of J/$\psi$ produced in p-Cu,
  p-U, O$_{\rm{16}}$-Cu, O$_{\rm{16}}$-U and S$_{\rm{32}}$-U collisions at
  200-GeV per nucleon},'' {\em Phys. Lett.}, vol.~B 262, pp.~362--368, 1991.

\bibitem{NA38_10}
C.~Baglin {\em et~al.}, ``{Transverse momentum of J/$\psi$ produced in oxygen
  uranium collisions at 200-GeV per nucleon},'' {\em Phys. Lett.}, vol.~B 251,
  pp.~465--471, 1990.

\bibitem{NA38_11}
M.~C. Abreu {\em et~al.}, ``{The production of J/$\psi$ in 200 GeV/A
  Oxygen-Uranium interactions },'' {\em Z. Phys.}, vol.~C 38, p.~117, 1988.

\bibitem{NA50_1}
M.~C. Abreu {\em et~al.}, ``{Anomalous J/$\psi$ suppression in Pb-Pb
  interactions at 158 GeV/c per nucleon},'' {\em Phys. Lett.}, vol.~B 410,
  pp.~337--343, 1997.

\bibitem{NA50_2}
M.~C. Abreu {\em et~al.}, ``{Observation of a threshold effect in the anomalous
  J/$\psi$ suppression},'' {\em Phys. Lett.}, vol.~B 450, pp.~456--466, 1999.

\bibitem{NA50_3}
M.~C. Abreu {\em et~al.}, ``{Dimuon and charm production in nucleus nucleus
  collisions at the CERN-SPS},'' {\em Eur. Phys. J.}, vol.~C 14, pp.~443--455,
  2000.

\bibitem{NA50_4}
M.~C. Abreu {\em et~al.}, ``{Evidence for deconfinement of quarks and gluons
  from the J/$\psi$ suppression pattern measured in Pb-Pb collisions at the
  CERN-SPS},'' {\em Phys. Lett.}, vol.~B 477, pp.~28--36, 2000.

\bibitem{NA50_5}
M.~C. Abreu {\em et~al.}, ``{Transverse momentum distributions of J/$\psi$,
  $\psi$', Drell- Yan and continuum dimuons produced in Pb-Pb interactions at
  the SPS},'' {\em Phys. Lett.}, vol.~B 499, pp.~85--96, 2001.

\bibitem{NA50_6}
M.~C. Abreu {\em et~al.}, ``{The dependence of the anomalous J/$\psi$
  suppression on the number of participant nucleons},'' {\em Phys. Lett.},
  vol.~B 521, pp.~195--203, 2001.

\bibitem{NA50_7}
B.~Alessandro {\em et~al.}, ``{Charmonia and Drell-Yan production in
  proton-nucleus collisions at the CERN SPS},'' {\em Phys. Lett.}, vol.~B 553,
  pp.~167--178, 2003.

\bibitem{NA50_8}
B.~Alessandro {\em et~al.}, ``{Charmonium production and nuclear absorption in
  p-A interactions at 450-GeV},'' {\em Eur. Phys. J.}, vol.~C 33, pp.~31--40,
  2004.

\bibitem{NA50_9}
B.~Alessandro {\em et~al.}, ``{A new measurement of J/$\psi$ suppression in
  Pb-Pb collisions at 158 GeV per nucleon},'' {\em Eur. Phys. J.}, vol.~C 39,
  pp.~335--345, 2005.

\bibitem{NA50_10}
B.~Alessandro {\em et~al.}, ``{Bottomonium and Drell-Yan production in p-A
  collisions at 450 GeV},'' {\em Phys. Lett.}, vol.~B 635, pp.~260--269, 2006.

\bibitem{NA50_11}
B.~Alessandro {\em et~al.}, ``{J/$\psi$ and $\psi$' production and their normal
  nuclear absorption in proton nucleus collisions at 400-GeV},'' {\em Eur.
  Phys. J.}, vol.~C 48, p.~329, 2006.

\bibitem{NA50_12}
B.~Alessandro {\em et~al.}, ``{$\psi$' production in Pb-Pb collisions at 158
  GeV/nucleon},'' {\em Eur. Phys. J.}, vol.~C 49, pp.~559--567, 2007.

\bibitem{NA50_13}
M.~Gonin {\em et~al.}, ``{Anomalous J/$\psi$ suppression in Pb-Pb collisions at
  158-A-GeV/c},'' {\em Nucl. Phys.}, vol.~A 610, pp.~404c--417c, 1996.

\bibitem{NA50_14}
M.~C. Abreu {\em et~al.}, ``{J/$\psi$ and Drell-Yan cross-sections in Pb-Pb
  interactions at 158 GeV/c per nucleon},'' {\em Phys. Lett.}, vol.~B 410,
  pp.~327--336, 1997.

\bibitem{NA50_15}
M.~C. Abreu {\em et~al.}, ``{Charmonium production in Pb-Pb interactions at
  158-GeV/c per nucleon},'' {\em Nucl. Phys.}, vol.~A 638, pp.~261--278, 1998.

\bibitem{NA50_16}
M.~C. Abreu {\em et~al.}, ``{Observation of a threshold effect in the anomalous
  J/$\psi$ suppression},'' {\em Phys. Lett.}, vol.~B 450, pp.~456--466, 1999.

\bibitem{NA51_1}
M.~C. Abreu {\em et~al.}, ``{J/$\psi$, $\psi$' and Drell-Yan production in pp
  and p-d interactions at 450-GeV/c},'' {\em Phys. Lett.}, vol.~B 438,
  pp.~35--40, 1998.

\bibitem{NA60_1}
R.~Arnaldi {\em et~al.}, ``{J/$\psi$ production in indium-indium collisions at
  158- GeV/nucleon},'' {\em Phys. Rev. Lett.}, vol.~99, p.~132302, 2007.

\bibitem{NA60_2}
R.~Arnaldi {\em et~al.}, ``{Evidence for the production of thermal-like muon
  pairs with masses above 1 GeV/$c^2$ in 158A GeV Indium-Indium Collisions},''
  {\em Eur. Phys. J.}, vol.~C 59, pp.~607--623, 2009.

\bibitem{E772_1}
D.~M. Alde {\em et~al.}, ``{Nuclear dependence of the production of $\Upsilon$
  resonances at 800-GeV},'' {\em Phys. Rev. Lett.}, vol.~66, pp.~2285--2288,
  1991.

\bibitem{E772_2}
D.~M. Alde {\em et~al.}, ``{The A-dependence of J/$\psi$ and $\psi$' production
  at 800-GeV/c},'' {\em Phys. Rev. Lett.}, vol.~66, pp.~133--136, 1991.

\bibitem{E772_3}
D.~M. Alde {\em et~al.}, ``{Nuclear dependence of the production of $\Upsilon$
  resonances at 800-GeV},'' {\em Phys. Rev. Lett.}, vol.~66, pp.~2285--2288,
  1991.

\bibitem{E772_4}
P.~L. McGaughey {\em et~al.}, ``{Cross-sections for the production of high mass
  muon pairs from 800-GeV proton bombardment of H-2},'' {\em Phys. Rev.},
  vol.~D 50, pp.~3038--3045, 1994.

\bibitem{E789_1}
M.~J. Leitch {\em et~al.}, ``{Nuclear dependence of neutral D meson production
  by 800- GeV/c protons},'' {\em Phys. Rev. Lett.}, vol.~72, pp.~2542--2545,
  1994.

\bibitem{E789_2}
M.~J. Leitch {\em et~al.}, ``{Nuclear dependence of neutral D meson production
  by 800- GeV/c protons},'' {\em Phys. Rev. Lett.}, vol.~72, pp.~2542--2545,
  1994.

\bibitem{E789_3}
C.~S. Mishra {\em et~al.}, ``{Search for the decay D$^0$ $\to$ $\mu^+
  \mu^-$},'' {\em Phys. Rev.}, vol.~D 50, pp.~9--12, 1994.

\bibitem{E789_4}
M.~H. Schub {\em et~al.}, ``{Measurement of J/$\psi$ and $\psi^\prime$
  production in 800-GeV/c proton - gold collisions},'' {\em Phys. Rev.}, vol.~D
  52, pp.~1307--1315, 1995.

\bibitem{E789_5}
M.~H. Schub {\em et~al.}, ``{Measurement of J/$\psi$ and $\psi$' production in
  800- GeV/c proton - gold collisions},'' {\em Phys. Rev.}, vol.~D 52, p.~1307,
  1995.

\bibitem{E789_6}
D.~M. Alde {\em et~al.}, ``{Nuclear dependence of dimuon production at 800-GeV.
  FNAL- 772 experiment},'' {\em Phys. Rev. Lett.}, vol.~64, pp.~2479--2482,
  1990.

\bibitem{E866_1}
M.~J. Leitch {\em et~al.}, ``{Measurement of J/$\psi$ and $\psi$' suppression
  in p-A collisions at 800-GeV/c},'' {\em Phys. Rev. Lett.}, vol.~84,
  pp.~3256--3260, 2000.

\bibitem{E866_2}
C.~N. Brown {\em et~al.}, ``{Observation of polarization in bottomonium
  production at $\sqrt{s}$ = 38.8-GeV},'' {\em Phys. Rev. Lett.}, vol.~86,
  pp.~2529--2532, 2001.

\bibitem{E866_3}
T.~H. Chang {\em et~al.}, ``{J/$\psi$ polarization in 800-GeV p-Cu
  interactions},'' {\em Phys. Rev. Lett.}, vol.~91, p.~211801, 2003.

\bibitem{E866_4}
L.~Y. Zhu {\em et~al.}, ``{Measurement of $\Upsilon$ Production for pp and p-d
  Interactions at 800 GeV/c},'' {\em Phys. Rev. Lett.}, vol.~100, p.~062301,
  2008.

\bibitem{HERA_1}
I.~Abt {\em et~al.}, ``{J/$\psi$ production via $\chi$(c) decays in 920-GeV pA
  interactions},'' {\em Phys. Lett.}, vol.~B 561, pp.~61--72, 2003.

\bibitem{HERA_2}
I.~Abt {\em et~al.}, ``{Measurement of the b$\bar{\rm{b}}$ production cross
  section in 920-GeV fixed-target proton nucleus collisions},'' {\em Eur. Phys.
  J.}, vol.~C 26, pp.~345--355, 2003.

\bibitem{HERA_3}
I.~Abt {\em et~al.}, ``{Search for the Flavor-Changing Neutral Current Decay
  $D^0 \to \mu^+\mu^-$ with the HERA-B Detector},'' {\em Phys. Lett.}, vol.~B
  596, pp.~173--183, 2004.

\bibitem{HERA_4}
I.~Abt {\em et~al.}, ``{Measurement of the $\Upsilon$ production cross-section
  in 920-GeV fixed-target proton-nucleus collisions},'' {\em Phys. Lett.},
  vol.~B 638, pp.~13--21, 2006.

\bibitem{HERA_5}
I.~Abt {\em et~al.}, ``{Measurement of the J/$\psi$ production cross section in
  920-GeV/c fixed-target proton-nucleus interactions},'' {\em Phys. Lett.},
  vol.~B 638, pp.~407--414, 2006.

\bibitem{HERA_6}
I.~Abt {\em et~al.}, ``{Improved measurement of the b$\bar{\rm{b}}$ production
  cross section in 920-GeV fixed-target proton nucleus collisions},'' {\em
  Phys. Rev.}, vol.~D 73, p.~052005, 2006.

\bibitem{HERA_7}
I.~Abt {\em et~al.}, ``{A Measurement of the $\psi^\prime$ to J/$\psi$
  production ratio in 920-GeV proton-nucleus interactions},'' {\em Eur. Phys.
  J.}, vol.~C 49, pp.~545--558, 2007.

\bibitem{HERA_8}
I.~Abt {\em et~al.}, ``{Bottom production cross section from double muonic
  decays of b-hadrons in 920-GeV proton nucleus collision},'' {\em Phys.
  Lett.}, vol.~B 650, pp.~103--110, 2007.

\bibitem{HERA_9}
I.~Abt {\em et~al.}, ``{Measurement of $D^0$,$ D^+$, $D_s^+$ and $D^{*+}$
  Production in Fixed Target 920 GeV Proton-Nucleus Collisions},'' {\em Eur.
  Phys. J.}, vol.~C 52, pp.~531--542, 2007.

\bibitem{HERA_10}
I.~Abt {\em et~al.}, ``{Production of the Charmonium States $\chi$(c1) and
  $\chi$(c2) in Proton Nucleus Interactions at $\sqrt{s}$ = 41.6-GeV},'' {\em
  Phys. Rev.}, vol.~D 79, p.~012001, 2009.

\bibitem{HERA_11}
I.~Abt {\em et~al.}, ``{Angular distributions of leptons from J/$\psi$ 's
  produced in 920 GeV fixed-target proton-nucleus collisions},'' {\em Eur.
  Phys. J.}, vol.~C 60, pp.~517--524, 2009.

\bibitem{HERA_12}
I.~Abt {\em et~al.}, ``{Kinematic distributions and nuclear effects of J/$\psi$
  production in 920 GeV fixed-target proton-nucleus collisions},'' {\em Eur.
  Phys. J.}, vol.~C 60, pp.~525--542, 2009.

\bibitem{PHENIX_1}
S.~S. Adler {\em et~al.}, ``{J/$\psi$ production in Au-Au collisions at
  $\sqrt{s_{NN}}$ = 200-GeV at the Relativistic Heavy Ion Collider},'' {\em
  Phys. Rev.}, vol.~C 69, p.~014901, 2004.

\bibitem{PHENIX_2}
S.~S. Adler {\em et~al.}, ``{J/$\psi$ production from proton proton collisions
  at $\sqrt{s}$ = 200-GeV},'' {\em Phys. Rev. Lett.}, vol.~92, p.~051802, 2004.

\bibitem{PHENIX_3}
S.~S. Adler {\em et~al.}, ``{J/$\psi$ production and nuclear effects for d-Au
  and p-p collisions at $\sqrt{s_{NN}}$ = 200-GeV},'' {\em Phys. Rev. Lett.},
  vol.~96, p.~012304, 2006.

\bibitem{PHENIX_4}
A.~Adare {\em et~al.}, ``{J/$\psi$ production vs centrality, transverse
  momentum, and rapidity in Au + Au collisions at $\sqrt{s_{NN}}$ = 200-
  GeV},'' {\em Phys. Rev. Lett.}, vol.~98, p.~232301, 2007.

\bibitem{PHENIX_5}
A.~Adare {\em et~al.}, ``{J/$\psi$ production versus transverse momentum and
  rapidity in pp collisions at $\sqrt{s}$ = 200-GeV},'' {\em Phys. Rev. Lett.},
  vol.~98, p.~232002, 2007.

\bibitem{PHENIX_6}
A.~Adare {\em et~al.}, ``{J/$\psi$ Production in $\sqrt{s_{\rm{NN}}} = 200$ GeV
  Cu-Cu Collisions},'' {\em Phys. Rev. Lett.}, vol.~101, p.~122301, 2008.

\bibitem{PHENIX_7}
A.~Adare {\em et~al.}, ``{Cold Nuclear Matter Effects on J/$\psi$ as
  Constrained by Deuteron-Gold Measurements at $\sqrt{s_{\rm{NN}}} = 200$
  GeV},'' {\em Phys. Rev.}, vol.~C 77, p.~024912, 2008.

\bibitem{PHENIX_8}
S.~Afanasiev {\em et~al.}, ``{Photoproduction of J/$\psi$ and of high mass e+e-
  in ultra- peripheral Au-Au collisions at $\sqrt{s_{\rm{NN}}} = 200$ GeV},''
  {\em Phys. Lett.}, vol.~B 679, pp.~321--329, 2009.

\bibitem{PHENIX_O_1}
S.~S. Adler {\em et~al.}, ``{Centrality dependence of charm production from
  single electrons measurement in Au-Au collisions at $\sqrt{s_{NN}}$ =
  200-GeV},'' {\em Phys. Rev. Lett.}, vol.~94, p.~082301, 2005.

\bibitem{PHENIX_O_2}
S.~S. Adler {\em et~al.}, ``{Nuclear modification of electron spectra and
  implications for heavy quark energy loss in Au-Au collisions at
  $\sqrt{s_{NN}}$ = 200-GeV},'' {\em Phys. Rev. Lett.}, vol.~96, p.~032301,
  2006.

\bibitem{PHENIX_O_3}
A.~Adare {\em et~al.}, ``{Energy Loss and Flow of Heavy Quarks in Au-Au
  Collisions at $\sqrt{s_{\rm{NN}}} = 200$ GeV},'' {\em Phys. Rev. Lett.},
  vol.~98, p.~172301, 2007.

\bibitem{PHENIX_O_4}
A.~Adare {\em et~al.}, ``{Measurement of high-p(T) single electrons from heavy-
  flavor decays in pp collisions at $\sqrt{s}$ = 200-GeV},'' {\em Phys. Rev.
  Lett.}, vol.~97, p.~252002, 2006.

\bibitem{PHENIX_O_5}
A.~Adare {\em et~al.}, ``{Measurement of Bottom versus Charm as a Function of
  Transverse Momentum with Electron-Hadron Correlations in pp Collisions at
  $\sqrt{s}=200$ GeV},'' {\em Phys. Rev. Lett.}, vol.~103, p.~082002, 2009.

\bibitem{PHENIX_O_6}
A.~Adare {\em et~al.}, ``{Dilepton mass spectra in pp collisions at $\sqrt{s}=
  200$ GeV and the contribution from open charm},'' {\em Phys. Lett.}, vol.~B
  670, pp.~313--320, 2009.

\bibitem{PHENIX_O_7}
S.~S. Adler {\em et~al.}, ``{Measurement of single muons at forward rapidity in
  pp collisions at $\sqrt{s}$ = 200-GeV and implications for charm
  production},'' {\em Phys. Rev.}, vol.~D 76, p.~092002, 2007.

\bibitem{PHENIX_O_8}
K.~Adcox {\em et~al.}, ``{Measurement of single electrons and implications for
  charm production in Au-Au collisions at $\sqrt{s_{NN}}$ = 130- GeV},'' {\em
  Phys. Rev. Lett.}, vol.~88, p.~192303, 2002.

\bibitem{STAR_1}
B.~I. Abelev {\em et~al.}, ``{J/$\psi$ production at high transverse momentum
  in pp and Cu-Cu collisions at $\sqrt{s_{\rm{NN}}}$=200GeV},'' {\em Phys.
  Rev.}, vol.~C 80, p.~041902, 2009.

\bibitem{STAR_O_1}
B.~I. Abelev {\em et~al.}, ``{Measurement of D* Mesons in Jets from pp
  Collisions at $\sqrt{s}$ = 200 GeV},'' {\em Phys. Rev.}, vol.~D 79,
  p.~112006, 2009.

\bibitem{STAR_O_2}
B.~I. Abelev {\em et~al.}, ``{Charmed hadron production at low transverse
  momentum in Au-Au collisions at RHIC},'' 2008.
\newblock arXiv: nucl-ex/0805.0364.

\bibitem{STAR_O_4}
J.~Adams {\em et~al.}, ``{Open charm yields in d-Au collisions at
  $\sqrt{s_{NN}}$ = 200-GeV},'' {\em Phys. Rev. Lett.}, vol.~94, p.~062301,
  2005.

\bibitem{UA1_1}
C.~Albajar {\em et~al.}, ``{Beauty production at the CERN p$\bar{\rm{p}}$
  collider},'' {\em Phys. Lett.}, vol.~B 256, pp.~121--128, 1991.

\bibitem{UA1_2}
C.~Albajar {\em et~al.}, ``{J/$\psi$ and $\psi$' production at the CERN
  p$\bar{\rm{p}}$ collider},'' {\em Phys. Lett.}, vol.~B 256, pp.~112--120,
  1991.

\bibitem{UA1_3}
C.~Albajar {\em et~al.}, ``{Measurement of the Bottom Quark Production
  Cross-Section in Proton - anti-Proton Collisions at $\sqrt{s} = 0.63$-
  TeV},'' {\em Phys. Lett.}, vol.~B 213, p.~405, 1988.

\bibitem{UA1_4}
C.~Albajar {\em et~al.}, ``{High transverse momentum J/$\psi$ production at the
  CERN proton/anti-proton collider},'' {\em Phys. Lett.}, vol.~B 200, p.~380,
  1988.

\bibitem{UA1_5}
C.~Albajar {\em et~al.}, ``{Low Mass Dimuon Production at the {CERN} Proton -
  Anti-Proton Collider},'' {\em Phys. Lett.}, vol.~B 209, p.~397, 1988.

\bibitem{UA1_6}
G.~Arnison {\em et~al.}, ``{Intermediate mass dimuon events at the CERN
  p$\bar{\rm{p}}$ collider at $\sqrt{s}$ = 540-GeV},'' {\em Phys. Lett.},
  vol.~B 155, p.~442, 1985.

\bibitem{UA6_1}
C.~Morel {\em et~al.}, ``{Measurement of the inclusive J/$\psi$ production
  cross- sections in $\bar{\rm{p}}$p and pp collisions at $\sqrt{s}$ = 24.3-
  GeV},'' {\em Phys. Lett.}, vol.~B 252, pp.~505--510, 1990.

\bibitem{CDF_1}
T.~Aaltonen {\em et~al.}, ``{Production of $\psi$(2S) Mesons in p$\bar{\rm{p}}$
  Collisions at 1.96 TeV},'' {\em Phys. Rev.}, vol.~D 80, p.~031103, 2009.

\bibitem{CDF_2}
A.~Abulencia {\em et~al.}, ``{Measurement of $\sigma$($\chi$(c2)B($\chi$(c2)
  $\to$ J/$\psi$ + $\gamma$) / $\sigma$($\chi$(c1)B($\chi$(c1) $\to$ J/$\psi$ +
  $\gamma$) in p$\bar{\rm{p}}$ collisions at $\sqrt{s}$ = 1.96-TeV},'' {\em
  Phys. Rev. Lett.}, vol.~98, p.~232001, 2007.

\bibitem{CDF_3}
A.~Abulencia {\em et~al.}, ``{Polarization of J/$\psi$ and $\psi$(2S) mesons
  produced in p$\bar{\rm{p}}$ collisions at $\sqrt{s} = 1.96$-TeV},'' {\em
  Phys. Rev. Lett.}, vol.~99, p.~132001, 2007.

\bibitem{CDF_4}
A.~Abulencia {\em et~al.}, ``{Measurement of the B$^+$ production cross section
  in p$\bar{\rm{p}}$ collisions at $\sqrt{s}$ = 1960-GeV},'' {\em Phys. Rev.},
  vol.~D 75, p.~012010, 2007.

\bibitem{CDF_5}
D.~Acosta {\em et~al.}, ``{Measurement of the J/$\psi$ meson and b-hadron
  production cross sections in p$\bar{\rm{p}}$ collisions at $\sqrt{s_{NN}}$ =
  1960-GeV},'' {\em Phys. Rev.}, vol.~D 71, p.~032001, 2005.

\bibitem{CDF_6}
D.~E. Acosta {\em et~al.}, ``{Measurement of prompt charm meson production
  cross- sections in p$\bar{\rm{p}}$ collisions at $\sqrt{s}$ = 1.96-TeV},''
  {\em Phys. Rev. Lett.}, vol.~91, p.~241804, 2003.

\bibitem{CDF_7}
T.~Aaltonen {\em et~al.}, ``{Observation of exclusive charmonium production and
  $\gamma$+$\gamma$ to $\mu^+\mu^-$ in p$\bar{\rm{p}}$ collisions at $\sqrt{s}
  = 1.96$ TeV},'' {\em Phys. Rev. Lett.}, vol.~102, p.~242001, 2009.

\bibitem{CDF_8}
D.~E. Acosta {\em et~al.}, ``{Cross-section for forward J/$\psi$ production in
  p$\bar{\rm{p}}$ collisions at S = 1.8-TeV},'' {\em Phys. Rev.}, vol.~D 66,
  p.~092001, 2002.

\bibitem{CDF_9}
D.~E. Acosta {\em et~al.}, ``{Branching ratio measurements of exclusive B$^+$
  decays to charmonium with the Collider Detector at Fermilab},'' {\em Phys.
  Rev.}, vol.~D 66, p.~052005, 2002.

\bibitem{CDF_10}
D.~E. Acosta {\em et~al.}, ``{$\Upsilon$ production and polarization in
  p$\bar{\rm{p}}$ collisions at $\sqrt{s} = 1.8$-TeV},'' {\em Phys. Rev.
  Lett.}, vol.~88, p.~161802, 2002.

\bibitem{CDF_11}
D.~E. Acosta {\em et~al.}, ``{Measurement of the B$^+$ total cross-section and
  B$^+$S differential cross-section d sigma / dp(T) in p$\bar{\rm{p}}$
  collisions at $\sqrt{s} = 1.8$-TeV},'' {\em Phys. Rev.}, vol.~D 65,
  p.~052005, 2002.

\bibitem{CDF_12}
A.~A. Affolder {\em et~al.}, ``{Observation of diffractive J/$\psi$ production
  at the Fermilab Tevatron},'' {\em Phys. Rev. Lett.}, vol.~87, p.~241802,
  2001.

\bibitem{CDF_13}
A.~A. Affolder {\em et~al.}, ``{Production of $\chi$(c1) and $\chi$(c2) in
  p$\bar{\rm{p}}$ collisions at $\sqrt{s}$ = 1.8-TeV},'' {\em Phys. Rev.
  Lett.}, vol.~86, pp.~3963--3968, 2001.

\bibitem{CDF_14}
A.~A. Affolder {\em et~al.}, ``{Measurement of J/$\psi$ and $\psi$(2S)
  polarization in p$\bar{\rm{p}}$ collisions at $\sqrt{s}$ = 1.8-TeV},'' {\em
  Phys. Rev. Lett.}, vol.~85, pp.~2886--2891, 2000.

\bibitem{CDF_15}
A.~A. Affolder {\em et~al.}, ``{Production of $\Upsilon$(1S) mesons from
  $\chi$(b) decays in p$\bar{\rm{p}}$ collisions at $\sqrt{s}$ = 1.8-TeV},''
  {\em Phys. Rev. Lett.}, vol.~84, pp.~2094--2099, 2000.

\bibitem{CDF_16}
F.~Abe {\em et~al.}, ``{J/$\psi$ and $\psi$(2S) production in p$\bar{\rm{p}}$
  collisions at $\sqrt{s}$ = 1.8-TeV},'' {\em Phys. Rev. Lett.}, vol.~79,
  pp.~572--577, 1997.

\bibitem{CDF_17}
F.~Abe {\em et~al.}, ``{Production of J/$\psi$ mesons from $\chi$(c) meson
  decays in p$\bar{\rm{p}}$ collisions at $\sqrt{s}$ = 1.8-TeV},'' {\em Phys.
  Rev. Lett.}, vol.~79, pp.~578--583, 1997.

\bibitem{CDF_18}
F.~Abe {\em et~al.}, ``{Measurement of the B meson differential cross-section,
  d sigma / d p(T), in p$\bar{\rm{p}}$ collisions at $\sqrt{s}$ = 1.8- TeV},''
  {\em Phys. Rev. Lett.}, vol.~75, pp.~1451--1455, 1995.

\bibitem{CDF_19}
F.~Abe {\em et~al.}, ``{$\Upsilon$ production in p$\bar{\rm{p}}$ collisions at
  $\sqrt{s}$ = 1.8-TeV},'' {\em Phys. Rev. Lett.}, vol.~75, p.~4358, 1995.

\bibitem{CDF_20}
F.~Abe {\em et~al.}, ``{Measurement of the bottom quark production
  cross-section using semileptonic decay electrons in p$\bar{\rm{p}}$
  collisions at $\sqrt{s}$ = 1.8-TeV},'' {\em Phys. Rev. Lett.}, vol.~71,
  pp.~500--504, 1993.

\bibitem{CDF_21}
F.~Abe {\em et~al.}, ``{Measurement of bottom quark production in 1.8-TeV
  p$\bar{\rm{p}}$ collisions using semileptonic decay muons},'' {\em Phys. Rev.
  Lett.}, vol.~71, pp.~2396--2400, 1993.

\bibitem{CDF_22}
F.~Abe {\em et~al.}, ``{Inclusive $\chi$(c) and b quark production in
  $\bar{\rm{p}}$p collisions at $\sqrt{s}$ = 1.8-TeV},'' {\em Phys. Rev.
  Lett.}, vol.~71, pp.~2537--2541, 1993.

\bibitem{CDF_23}
F.~Abe {\em et~al.}, ``{A Measurement of the B meson and b quark cross-sections
  at $\sqrt{s}$ = 1.8-TeV using the exclusive decay B$^+-$ $\to$ J/$\psi$
  K$^+-$},'' {\em Phys. Rev. Lett.}, vol.~68, pp.~3403--3407, 1992.

\bibitem{CDF_24}
F.~Abe {\em et~al.}, ``{Inclusive J/$\psi$, $\psi$(2S) and b quark production
  in $\bar{\rm{p}}$p collisions at $\sqrt{s}$ = 1.8-TeV},'' {\em Phys. Rev.
  Lett.}, vol.~69, pp.~3704--3708, 1992.

\bibitem{D0_1}
S.~Abachi {\em et~al.}, ``{Measurement of the $\Upsilon$ cross-section at D0
  using dimuons},''
\newblock Submitted to International Europhysics Conference on High Energy
  Physics (HEP 95), Brussels, Belgium, 27 Jul - 2 Aug 1995.

\bibitem{D0_2}
S.~Abachi {\em et~al.}, ``{Inclusive muon and B quark production cross-sections
  in p$\bar{\rm{p}}$ collisions at $\sqrt{s}$ = 1.8-TeV},''
\newblock To be published in the proceedings of International Europhysics
  Conference on High Energy Physics (HEP 95), Brussels, Belgium, 27 Jul - 2 Aug
  1995.

\bibitem{D0_3}
K.~A. Bazizi, ``{Inclusive B quark and heavy quarkonium production at D0},''
  {\em AIP Conf. Proc.}, vol.~357, pp.~105--119, 1996.

\bibitem{D0_4}
S.~Abachi {\em et~al.}, ``{J/$\psi$ production in p$\bar{\rm{p}}$ collisions at
  $\sqrt{s}$ = 1.8-TeV},'' {\em Phys. Lett.}, vol.~B 370, pp.~239--248, 1996.

\bibitem{D0_5}
B.~Abbott {\em et~al.}, ``{Small angle J/$\psi$ production in p$\bar{\rm{p}}$
  collisions at $\sqrt{s}$ = 1.8-TeV},'' {\em Phys. Rev. Lett.}, vol.~82,
  pp.~35--40, 1999.

\bibitem{D0_6}
B.~Abbott {\em et~al.}, ``{The b$\bar{\rm{b}}$ production cross-section and
  angular correlations in p$\bar{\rm{p}}$ collisions at $\sqrt{s}$ = 1.8-
  TeV},'' {\em Phys. Lett.}, vol.~B 487, pp.~264--272, 2000.

\bibitem{D0_7}
B.~Abbott {\em et~al.}, ``{Small angle muon and bottom quark production in
  p$\bar{\rm{p}}$ collisions at $\sqrt{s}$ = 1.8-TeV},'' {\em Phys. Rev.
  Lett.}, vol.~84, pp.~5478--5483, 2000.

\bibitem{D0_8}
B.~Abbott {\em et~al.}, ``{Cross-section for b-jet production in
  $\bar{\rm{p}}$p collisions at $\sqrt{s}$ = 1.8-TeV},'' {\em Phys. Rev.
  Lett.}, vol.~85, pp.~5068--5073, 2000.

\bibitem{D0_9}
V.~M. Abazov {\em et~al.}, ``{Measurement of inclusive differential cross
  sections for $\Upsilon$(1S) production in p$\bar{\rm{p}}$ collisions at
  $\sqrt{s}$ = 1.96-TeV},'' {\em Phys. Rev. Lett.}, vol.~94, p.~232001, 2005.

\bibitem{ALICE_1}
K.~Aamodt {\em et~al.}, ``{Rapidity and transverse momentum dependence of
  inclusive J/$\psi$ production in pp collisions at $\sqrt{s} = 7$ TeV},'' {\em
  Phys. Lett.}, vol.~B 704, pp.~442--455, 2011.
\newblock Erratum-ibid. \textit{Phys. Lett.}, vol. B 718, p. 692, 2012.

\bibitem{ALICE_2}
B.~Abelev {\em et~al.}, ``{Measurement of charm production at central rapidity
  in proton-proton collisions at $\sqrt{s} = 7$ TeV},'' {\em JHEP}, vol.~1201,
  p.~128, 2012.

\bibitem{ALICE_3}
B.~Abelev {\em et~al.}, ``{J/$\psi$ polarization in pp collisions at
  $\sqrt{s}=7$ TeV},'' {\em Phys. Rev. Lett.}, vol.~108, p.~082001, 2012.

\bibitem{ALICE_4}
B.~Abelev {\em et~al.}, ``{Heavy flavour decay muon production at forward
  rapidity in proton--proton collisions at $\sqrt{s} = 7$ TeV},'' {\em Phys.
  Lett.}, vol.~B 708, pp.~265--275, 2012.

\bibitem{ALICE_5}
B.~Abelev {\em et~al.}, ``{J/$\psi$ suppression at forward rapidity in Pb-Pb
  collisions at $\sqrt{s_{NN}}=2.76$ TeV},'' {\em Phys. Rev. Lett.}, vol.~109,
  p.~072301, 2012.

\bibitem{ALICE_6}
B.~Abelev {\em et~al.}, ``{J/$\psi$ Production as a Function of Charged
  Particle Multiplicity in pp Collisions at $\sqrt{s} = 7$ TeV},'' {\em Phys.
  Lett.}, vol.~B 712, pp.~165--175, 2012.

\bibitem{ALICE_7}
B.~Abelev {\em et~al.}, ``{Suppression of high transverse momentum D mesons in
  central Pb-Pb collisions at $\sqrt{s_{NN}}=2.76$ TeV},'' {\em JHEP},
  vol.~1209, p.~112, 2012.

\bibitem{ALICE_8}
B.~Abelev {\em et~al.}, ``{Inclusive J/$\psi$ production in pp collisions at
  $\sqrt{s} = 2.76$ TeV},'' {\em Phys. Lett.}, vol.~B 718, pp.~295--306, 2012.

\bibitem{ATLAS_1}
G.~Aad {\em et~al.}, ``{Measurement of the differential cross-sections of
  inclusive, prompt and non-prompt $J/\psi$ production in proton-proton
  collisions at $\sqrt{s}=7$ TeV},'' {\em Nucl. Phys.}, vol.~B 850,
  pp.~387--444, 2011.

\bibitem{ATLAS_2}
G.~Aad {\em et~al.}, ``{Measurement of the differential cross-sections of
  inclusive, prompt and non-prompt J/$\psi$ production in proton-proton
  collisions at $\sqrt{s}=7$ TeV},'' {\em Nucl. Phys.}, vol.~B 850,
  pp.~387--444, 2011.

\bibitem{ATLAS_3}
G.~Aad {\em et~al.}, ``{Measurement of the $\Upsilon$(1S) Production
  Cross-Section in pp Collisions at $\sqrt{s}=7$ TeV in ATLAS},'' {\em Phys.
  Lett.}, vol.~B 705, pp.~9--27, 2011.

\bibitem{ATLAS_4}
G.~Aad {\em et~al.}, ``{Search for dilepton resonances in pp collisions at
  $\sqrt{s}=7$ TeV with the ATLAS detector},'' {\em Phys. Rev. Lett.},
  vol.~107, p.~272002, 2011.

\bibitem{ATLAS_5}
G.~Aad {\em et~al.}, ``{Measurements of the electron and muon inclusive
  cross-sections in proton-proton collisions at $\sqrt{s}=7$ TeV with the ATLAS
  detector},'' {\em Phys. Lett.}, vol.~B 707, pp.~438--458, 2012.

\bibitem{ATLAS_6}
G.~Aad {\em et~al.}, ``{Measurement of $D^{*+/-}$ meson production in jets from
  pp collisions at $\sqrt{s}$ = 7 TeV with the ATLAS detector},'' {\em Phys.
  Rev.}, vol.~D 85, p.~052005, 2012.

\bibitem{ATLAS_7}
G.~Aad {\em et~al.}, ``{Observation of a new $\chi$(b) state in radiative
  transitions to $\Upsilon$(1S) and $\Upsilon$(2S) at ATLAS},'' {\em Phys. Rev.
  Lett.}, vol.~108, p.~152001, 2012.

\bibitem{ATLAS_8}
G.~Aad {\em et~al.}, ``{Measurement of $\Upsilon$ production in 7 TeV pp
  collisions at ATLAS},'' 2012.
\newblock CERN-PH-EP-2012-295.

\bibitem{CMS_1}
V.~Khachatryan {\em et~al.}, ``{Prompt and non-prompt J/$\psi$ production in pp
  collisions at $\sqrt{s} = 7$ TeV},'' {\em Eur. Phys. J.}, vol.~C 71, p.~1575,
  2011.

\bibitem{CMS_2}
V.~Khachatryan {\em et~al.}, ``{Measurement of the Inclusive $\Upsilon$
  production cross section in pp collisions at $\sqrt{s}=7$ TeV},'' {\em Phys.
  Rev.}, vol.~D 83, p.~112004, 2011.

\bibitem{CMS_3}
V.~Khachatryan {\em et~al.}, ``{Measurement of the B$^+$ Production Cross
  Section in pp Collisions at $\sqrt{s} = 7$ TeV},'' {\em Phys. Rev. Lett.},
  vol.~106, p.~112001, 2011.

\bibitem{CMS_4}
V.~Khachatryan {\em et~al.}, ``{Inclusive b-hadron production cross section
  with muons in pp collisions at $\sqrt{s} = 7$ TeV},'' {\em JHEP}, vol.~03,
  p.~090, 2011.

\bibitem{CMS_5}
V.~Khachatryan {\em et~al.}, ``{Measurement of B$\bar{\rm{B}}$ Angular
  Correlations based on Secondary Vertex Reconstruction at $\sqrt{s}=7$ TeV},''
  {\em JHEP}, vol.~03, p.~136, 2011.

\bibitem{CMS_6}
S.~Chatrchyan {\em et~al.}, ``{Measurement of the B0 production cross section
  in pp Collisions at $\sqrt{s} = 7$ TeV},'' {\em Phys. Rev. Lett.}, vol.~106,
  p.~252001, 2011.

\bibitem{CMS_7}
S.~Chatrchyan {\em et~al.}, ``{Measurement of the Strange B Meson Production
  Cross Section with J/$\psi$ $\phi$ Decays in pp Collisions at $\sqrt{s} = 7$
  TeV},'' {\em Phys. Rev.}, vol.~D 84, p.~052008, 2011.

\bibitem{CMS_8}
S.~Chatrchyan {\em et~al.}, ``{Search for B(s) and B to dimuon decays in pp
  collisions at 7 TeV},'' {\em Phys. Rev. Lett.}, vol.~107, p.~191802, 2011.

\bibitem{CMS_9}
S.~Chatrchyan {\em et~al.}, ``{J/$\psi$ and $\psi$(2S) production in pp
  collisions at $\sqrt{s}=7$ TeV},'' {\em JHEP}, vol.~1202, p.~011, 2012.

\bibitem{CMS_10}
S.~Chatrchyan {\em et~al.}, ``{Inclusive $b$-jet production in pp collisions at
  $\sqrt{s}=7$ TeV},'' {\em JHEP}, vol.~1204, p.~084, 2012.

\bibitem{CMS_11}
S.~Chatrchyan {\em et~al.}, ``{Measurement of the cross section for production
  of $b b^-$ bar $X$, decaying to muons in pp collisions at $\sqrt{s}=7$
  TeV},'' {\em JHEP}, vol.~1206, p.~110, 2012.

\bibitem{CMS_12}
S.~Chatrchyan {\em et~al.}, ``{Search for $B^0_s \to \mu^+ \mu^-$ and $B^0 \to
  \mu^+ \mu^-$ decays},'' {\em JHEP}, vol.~1204, p.~033, 2012.

\bibitem{CMS_13}
S.~Chatrchyan {\em et~al.}, ``{Measurement of the $\Upsilon$(1S),
  $\Upsilon$(2S) and $\Upsilon$(3S) polarizations in pp collisions at
  $\sqrt{s}=7$ TeV},'' {\em Phys. Rev. Lett.}, 2012.

\bibitem{LHCb_1}
R.~Aaij {\em et~al.}, ``{Measurement of $\sigma(pp \to b \bar{b} X)$ at
  $\sqrt{s}=7~\rm{TeV}$ in the forward region},'' {\em Phys. Lett.}, vol.~B
  694, pp.~209--216, 2010.

\bibitem{LHCb_2}
R.~Aaij {\em et~al.}, ``{First observation of $B^0_s \to J/\psi f_0(980)$
  decays},'' {\em Phys. Lett.}, vol.~B 698, pp.~115--122, 2011.

\bibitem{LHCb_3}
R.~Aaij {\em et~al.}, ``{Measurement of J/$\psi$ production in pp collisions at
  $\sqrt{s}=7~\rm{TeV}$},'' {\em Eur. Phys. J.}, vol.~C 71, p.~1645, 2011.

\bibitem{LHCb_4}
R.~Aaij {\em et~al.}, ``{Observation of J/$\psi$ pair production in pp
  collisions at $\sqrt{s}=7$ TeV},'' {\em Phys. Lett.}, vol.~B 707, pp.~52--59,
  2012.

\bibitem{LHCb_5}
R.~Aaij {\em et~al.}, ``{Measurement of $b$-hadron production fractions in
  $7~\rm{TeV}$ pp collisions},'' {\em Phys. Rev.}, vol.~D 85, p.~032008, 2012.

\bibitem{LHCb_6}
R.~Aaij {\em et~al.}, ``{Measurement of the $B^\pm$ production cross-section in
  pp collisions at $\sqrt{s}=7$ TeV},'' {\em JHEP}, vol.~1204, p.~093, 2012.

\bibitem{LHCb_7}
R.~Aaij {\em et~al.}, ``{Measurement of $\Upsilon$ production in pp collisions
  at $\sqrt{s}=7$ TeV},'' {\em Eur. Phys. J.}, vol.~C 72, p.~2025, 2012.

\bibitem{LHCb_8}
R.~Aaij {\em et~al.}, ``{Measurement of the cross-section ratio $\sigma (
  \chi(c_2) ) / \sigma(\chi(c_1))$ for prompt $\chi$(c) production at
  $\sqrt{s}=7$ TeV},'' {\em Phys. Lett.}, vol.~B 714, pp.~215--223, 2012.

\bibitem{LHCb_9}
R.~Aaij {\em et~al.}, ``{Measurement of the ratio of prompt $\chi$(c) to
  J/$\psi$ production in pp collisions at $\sqrt{s}=7$ TeV},'' {\em Phys.
  Lett.}, vol.~B 718, pp.~431--440, 2012.

\bibitem{Inspires_webpage}
http://inspirehep.net/.

\end{thebibliography}

\end{document}